\documentclass[prd, nofootinbib]{revtex4}
\usepackage{amssymb,amsmath,microtype,url}
\usepackage{graphicx}
\usepackage{latexsym}
\usepackage[normalem]{ulem}
\usepackage{appendix}

\def\bea{\begin{eqnarray}}
\def\eea{\end{eqnarray}}

\def\ie{{\frenchspacing\it i.e.}}

\def\be{\begin{equation}}
\def\ee{\end{equation}}

\usepackage{color}

\begin{document}

\title{Probing Primordial Gravitational Waves: Ali CMB Polarization Telescope\footnote{Invited Paper for $National ~Science ~Review$}}

\author{Hong Li$^{1}$, Si-Yu Li$^{1}$, Yang Liu$^{2,3}$, Yong-Ping Li$^{2,3}$, Yifu Cai$^{4}$, Mingzhe Li$^{5}$, Gong-Bo Zhao$^{6,7}$, Cong-Zhan Liu$^{1}$, Zheng-Wei Li$^{1}$, He Xu$^{1}$, Di Wu$^{1}$, Yong-Jie Zhang$^{1}$, Zu-Hui Fan$^{8}$,  Yong-Qiang Yao$^{6}$, Chao-Lin Kuo $^{9}$, Fang-Jun Lu$^{1}$ and Xinmin Zhang$^{2,3}$}

\affiliation{$^1$Key Laboratory of Particle Astrophysics,  Institute of High Energy Physics (IHEP), Chinese Academy of Sciences, 19B Yuquan Road, Shijingshan District, Beijing 100049, China}
\affiliation{$^2$Theoretical Physics Division, Institute of High Energy Physics (IHEP), Chinese Academy of Sciences, 19B Yuquan Road, Shijingshan District, Beijing 100049, China}
\affiliation{$^3$University of Chinese Academy of Sciences, Beijing, China}
\affiliation{$^4$CAS Key Laboratory for Researches in Galaxies and Cosmology, Department of Astronomy, University of Science and Technology of China, Hefei, Anhui 230026, China}
\affiliation{$^5$Interdisciplinary Center for Theoretical Study, University of Science and Technology of China, Hefei, Anhui 230026, China}
\affiliation{$^6$National Astronomical Observatories, Chinese Academy of Science, Jia 20, Datun Road, Chaoyang District, Beijing 100012, P.R.China}
\affiliation{$^7$Institute of Cosmology \& Gravitation, University of Portsmouth, Dennis Sciama Building, Portsmouth, PO1 3FX, UK}
\affiliation{$^8$Department of Astronomy, School of Physics, Peking University, Beijing 100871, China}
\affiliation{$^9$Physics Department, Stanford University, 385 Via Pueblo Mall, Stanford, CA 94305, USA}

\begin{abstract}
In this paper, we will give a general introduction to the project of Ali CMB Polarization Telescope (AliCPT), which is a Sino-US joint project led by the Institute of
High Energy Physics (IHEP) and involves many different institutes in China. It is the first ground-based cosmic microwave background (CMB) polarization experiment in China
and an integral part of China's {\it Gravitational Waves Program}. The main scientific goal of AliCPT project is to probe
the primordial gravitational waves (PGWs) originated from the very early Universe.

The AliCPT project includes two stages.
The first stage referred to as AliCPT-1, is to build a telescope in the Ali region of Tibet with an altitude of 5,250 meters.
Once completed, it will be the worldwide highest ground-based CMB observatory and open a new window for probing PGWs in northern hemisphere. AliCPT-1 telescope is designed to have about 7,000 TES detectors at 95GHz and 150GHz.
The second stage is to have a more sensitive telescope (AliCPT-2)  with the number of detectors more than 20,000.

Our simulations show that AliCPT will improve the current constraint on the tensor-to-scalar ratio $r$ by one order of magnitude with 3 years' observation. Besides the PGWs, the AliCPT will also enable a precise measurement on the CMB rotation angle and provide a precise test on the CPT symmetry. We show 3 years' observation will improve the current limit by two order of magnitude.

\end{abstract}
\maketitle

\section{Introduction}

Searching for gravitational waves (GWs) has long been the cornerstone of cosmology and astrophysics since Einstein
proposed the General Relativity (GR) in early 20th century. GWs are thought to be the last piece of the theoretical predictions of GR.
With long-lasting efforts, LIGO (Laser Interferometer Gravitational-Wave Observatory) collaboration in 2016
announced the first detection of GWs with the signals coming from two merging black holes with mass of tens of solar masses \cite{Abbott:2016blz}. Since then, LIGO and Virgo have announced other three events of black hole \cite{Abbott:2016nmj, Abbott:2017vtc, Abbott:2017oio} and one event of neutron stars GWs \cite{LIGO:Neutron}.
These achievements make the GWs study to enter a new era and were awarded the Nobel Prize in Physics in 2017.

Different from the GWs detected by LIGO and Virgo, the PGWs arise from quantum fluctuations and carry important information about the very early Universe, for example, the physics of inflation, bouncing and emergent Universe.
So far the most effective way to probe PGWs is to measure the B-mode polarization of CMB.

The CMB photons are relics left after the Big Bang. Its first detection half a century ago pioneered the study of cosmology. For the recent 20 to 30 years, CMB observations have developed rapidly, leading us into the precision cosmology era.
However, the CMB B-mode polarization induced by the tensor fluctuations generated in the early Universe, i.e., the PGWs, still have not been detected conclusively\footnote{ BICEP2 collaboration\cite{BICEP2} announced in 2014 the detection of B-mode, which however turns out to be the dust emissions dominate, not the signal of PGWs\cite{Planckdust}.}. This has become a key scientific goal of CMB observations in recent years.
In addition, CMB B-modes provide us an important test on fundamental physics, such as the CPT symmetry.

At present, major ground-based CMB experiments are in southern hemisphere, for example, the Atacama Cosmology Telescope (ACT) and POLARBEAR/Simons Array in Chile, and the South Pole Telescope (SPT) and BICEP at the South Pole. High precision experiments in the northern hemisphere are critically needed to achieve a full sky coverage.

In 2014, IHEP cosmology team proposed a CMB experiment in Ali of Tibet, aiming to search for the PGWs in northern hemisphere.
In this paper we provide a general introduction to the AliCPT project.
In section \ref{section2}, we introduce the atmospheric conditions, sky coverage and infrastructure of AliCPT. In section \ref{section3}, we present the main scientific goals of AliCPT. Section \ref{section4} is our summary.

\section{Overview of the AliCPT site}
\label{section2}

In this section, we describe the atmospheric conditions, sky coverage and the infrastructure of AliCPT site.

\begin{itemize}

\item\textbf{Atmospheric conditions:}
For ground-based CMB telescopes, the atmosphere is a big concern. The absorption and emission at millimeter/sub-millimeter band by the air molecules reduce the significance of signals. Among all components of the air, water vapor plays a crucial role due to its strong absorption/emission and heavy time variation. Usually, the Precipitable Water Vapor (PWV) is used as a conventional parameter to characterize the amount of water vapor, which is defined as the overall depth of water in a column of the atmosphere above the ground.  CMB signals, especially the polarization signals, are extremely weak,  so observation requires the air of the site to be thin, dry and stable. In the upper panel of Figure \ref{fig:globaldis}, we show the global distribution of the mean values of PWV over the past 6 years (2011.7-2017.7). It's obvious that only four regions have the lowest PWV on earth, including the Antarctic, Atacama Desert, Greenland and the high Tibetan Plateau.

\begin{figure}
\begin{center}
\includegraphics[scale=0.35]{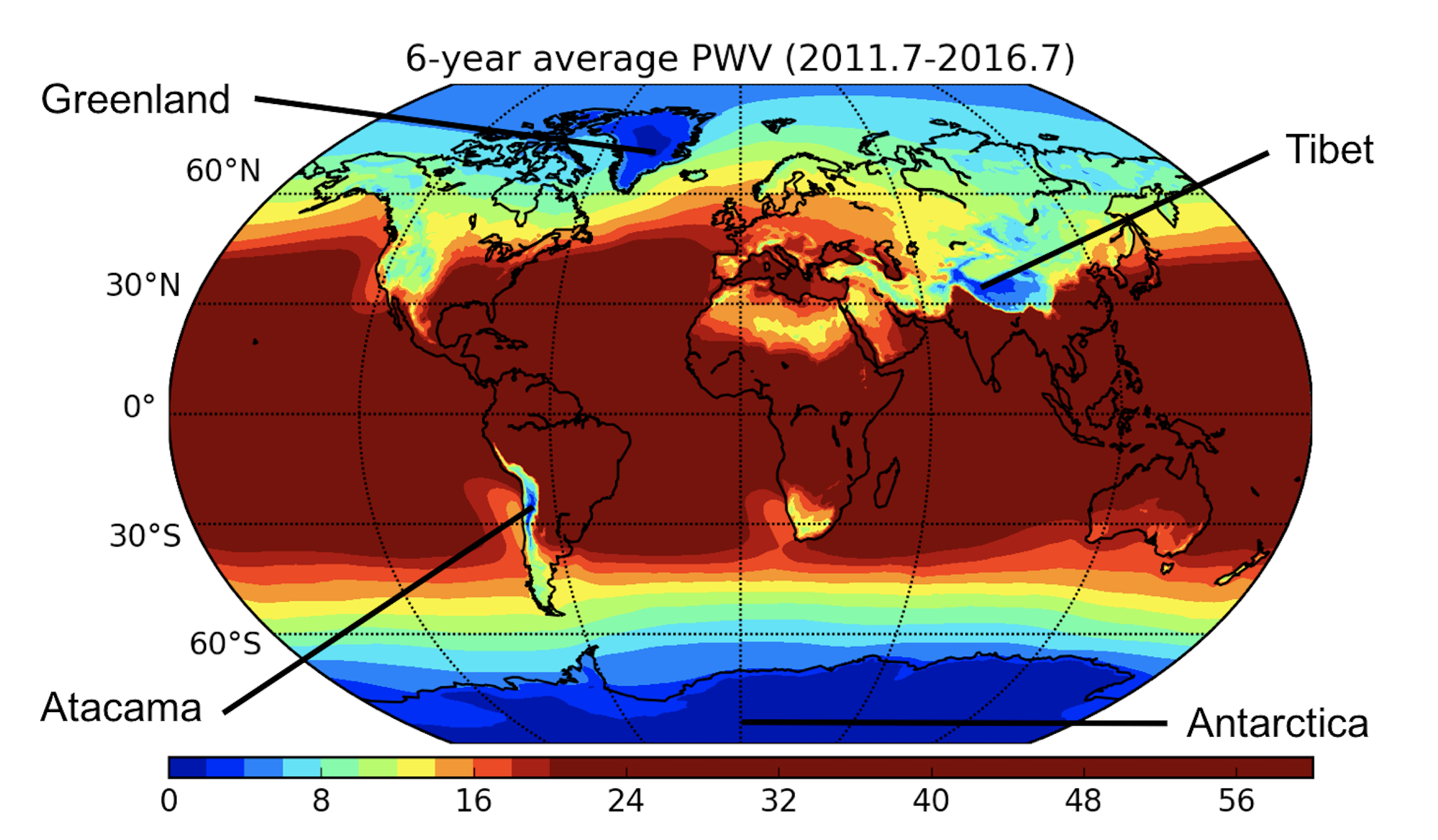}
\includegraphics[width=0.8\textwidth]{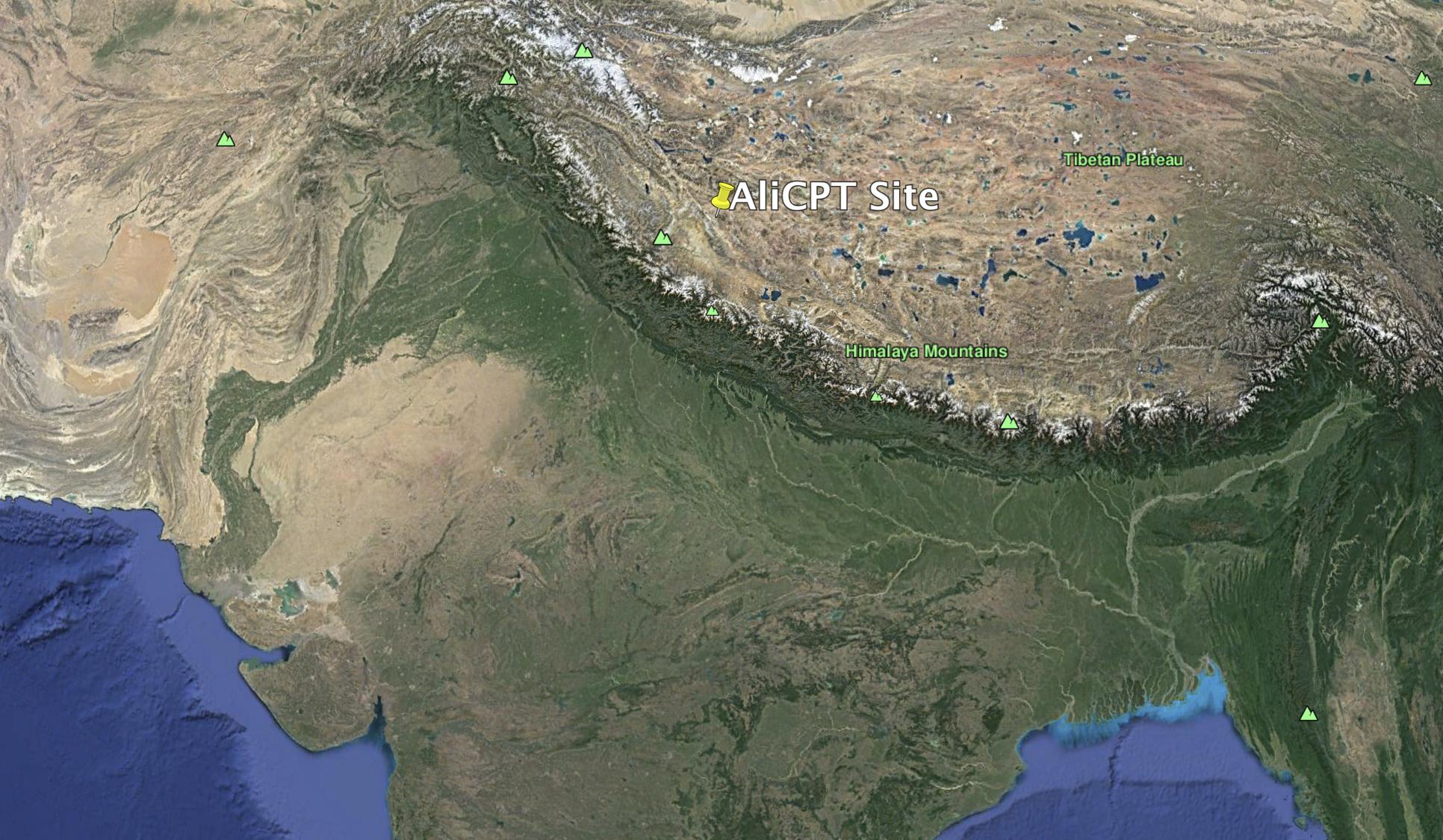}
\caption{Global distribution of mean PWV over 6 years (2011.7-2017.7) obtained with MERRA-2 data. The color bar is in the unit of millimeter (upper). The location of AliCPT site and the terrain around. The wet air from the Indian Ocean is hugely reduced by the Himalaya Mountain(lower).}
\label{fig:globaldis}
\end{center}
\end{figure}

\begin{figure}
\begin{center}
\includegraphics[scale=0.4]{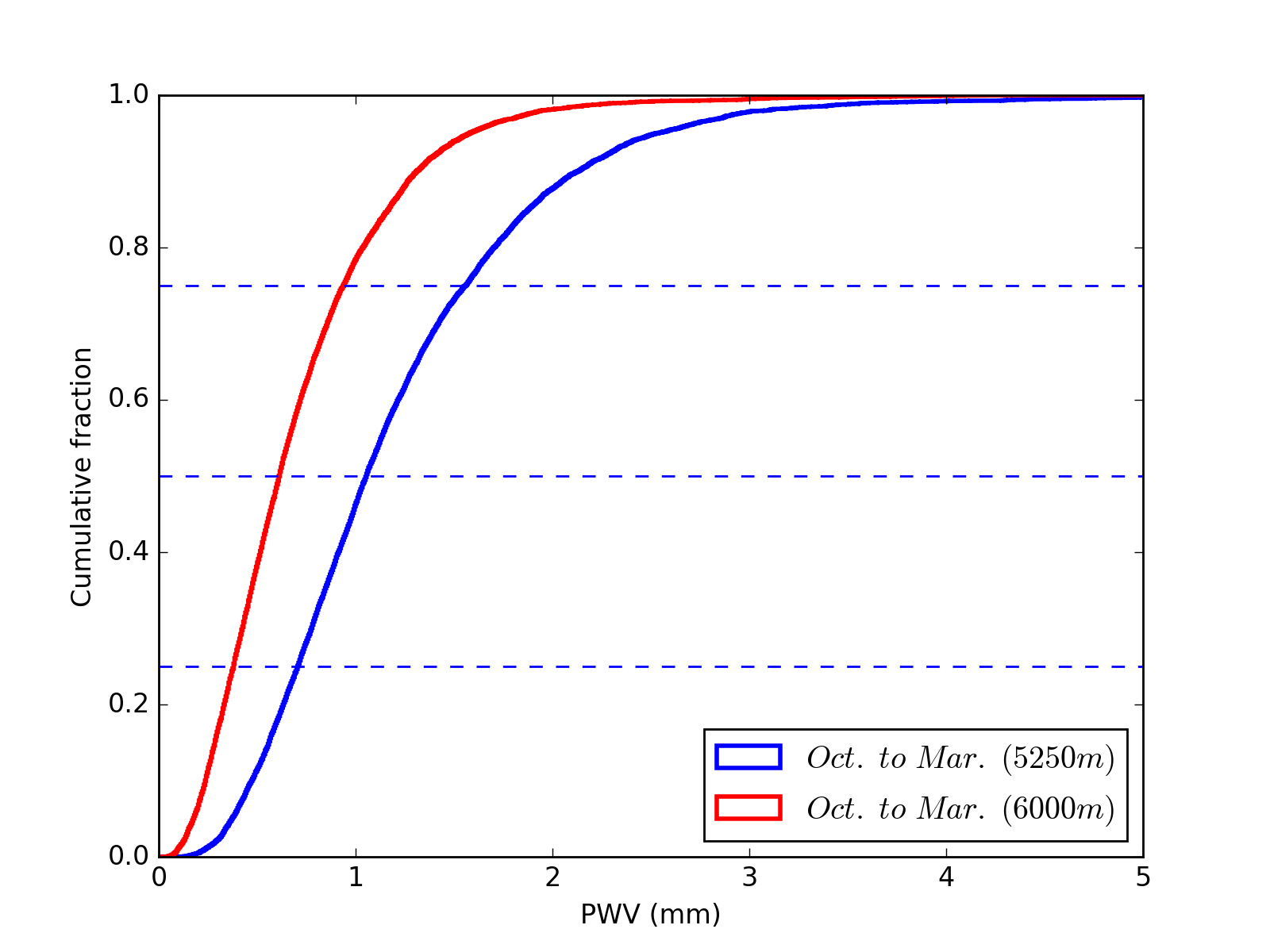}
\includegraphics[scale=0.4]{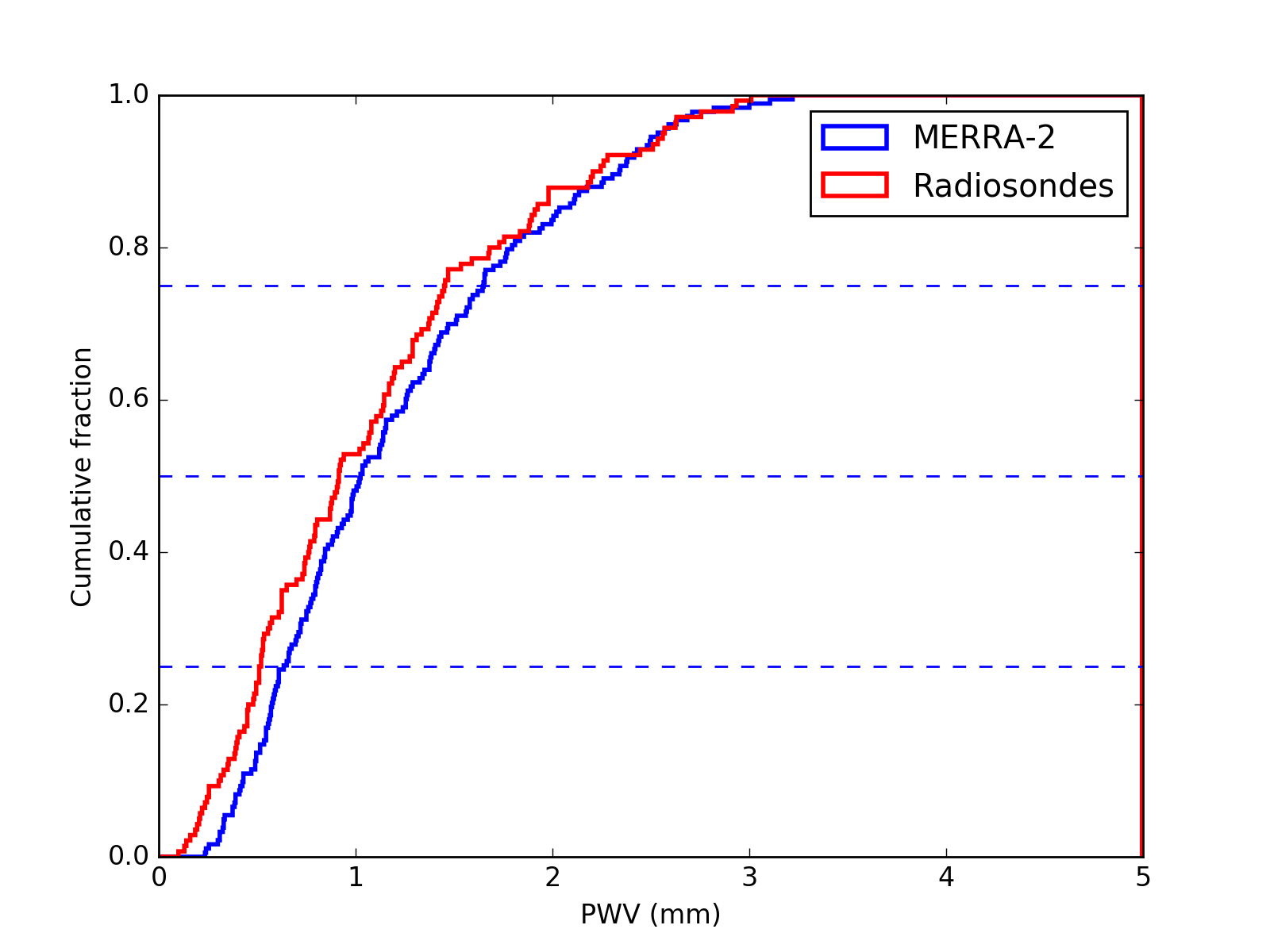}
\caption{The cumulative distribution of PWV for sites at 5,250m and 6,000m over observing season obtained with MERRA-2 data (left). The comparison between MERRA-2 and radiosondes datasets at 5250m (right). Results are taken from \cite{Ping}.}
\label{fig:culmulative}
\end{center}
\end{figure}

Current AliCPT site is located at a 5,250m high peak of the Gangdise Mountain. Another site nearby with altitude of about 6,000m has also been considered for the forthcoming project. The Himalayas is to its southwest and runs from northwest to southeast, separating Ali area from the Indian subcontinent as well as the Indian Ocean, which is shown in the lower panel of Figure \ref{fig:globaldis}.  So the wet air from the Indian Ocean is largely reduced. These make air of AliCPT site to be thin and dry enough around Winter. In \cite{Ping}, we quantitatively analyzed the atmospheric conditions of the site using radiosondes data from the local weather station as well as MERRA-2 reanalysis data from NASA/GMAO. The results show that PWV of Ali have a very strong seasonal variation and the median PWV of the observing season (October to March) is about 1mm (1.07mm for MERRA-2, 0.92mm for radiosondes), which is excellent to observation at 95/150GHz. In Figure \ref{fig:culmulative} we plot, as a function of PWV, the fraction of time during which the water vapor condition is better than a specific PWV value. References \cite{Ye, Kuo} have also done the atmospherical conditions evaluation for Ali region.

\item\textbf{Sky coverage:}
 The AliCPT site is located at geographical coordination ($80^{\circ}01'50''E$ , $32^{\circ}18'38''N$). With the rotation of the earth and the mid-latitude location, AliCPT is able to cover the whole northern sky as well as the low latitude part of the southern sky, and the overall observable fraction is about 70 percent. We show this in Figure \ref{fig:cover} as the region above the black dashed line. In our calculation we choose the instrumental parameters to be a $45^\circ$ lowest elevation for the mount and a 30$^\circ$ field of view (FOV). The overlap in the low latitude region of observable sky of Ali and Atacama makes it convenient to do the cross-check and cross-correlation studies.
Known as the northern hole, the lowest foreground contaminated region in the northern galactic hemisphere is also within the observable sky of AliCPT, which is extremely important for a CMB B-mode polarization and PGWs aimed project. We show the target fields of AliCPT in Figure \ref{fig:cover}. TN1 and TN2 within the black solid lines are target fields in the northern galactic hemisphere and TS in the southern galactic hemisphere. TND with the lowest dust intensity is chosen for deeper survey.

So the sky coverage (both observable and low foreground contamination sky) of AliCPT is complement to that of experiments in Antarctic and Atacama. Together with the southern cleanest sky region that covered by southern projects such as BICEP and Simons Array, AliCPT will increase the chance to find the B mode and PGWs. For more details about sky coverage, we refer to \cite{Ping} .

\begin{figure}
\begin{center}
\includegraphics[scale=0.5]{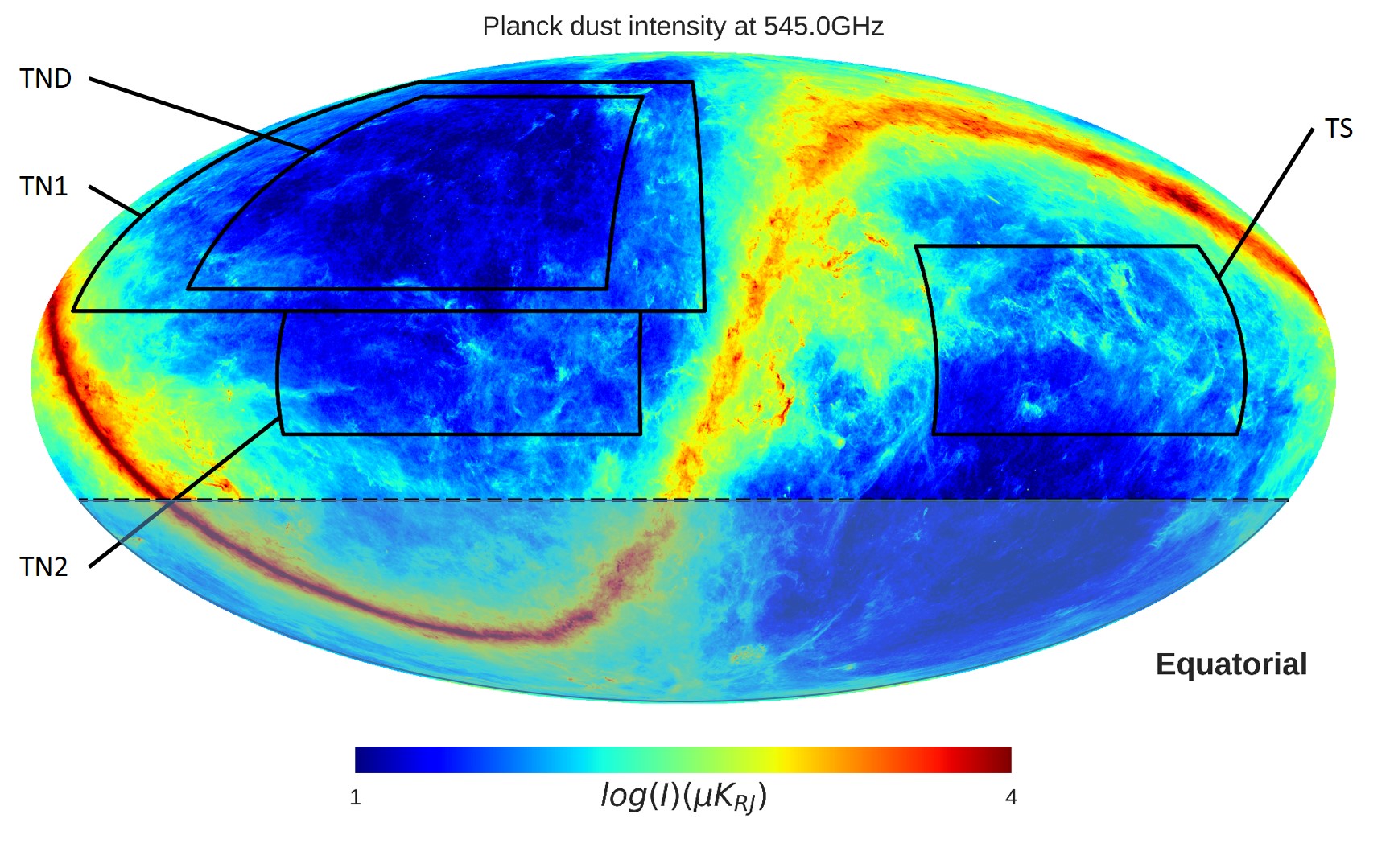}
\caption{Sky coverage of AliCPT. The observable sky is above the black dashed line, the target field is shown as TN1 and TN2 in the northern galactic hemisphere and TS in the southern galactic hemisphere. TND is for deeper survey. }
\label{fig:cover}
\end{center}
\end{figure}

\item\textbf{Infrastructure: }
On the transportation, the National Highway 219 is right next to the AliCPT site. Moreover, the Ngari Gunsa Airport is only about half an hour driving distance to the site and has a daily commercial flight to Lhasa, the capital of Tibet. The largest settlement of Ali area, Shiquanhe town is also very close to the site and it takes only 30 minutes driving to get there.

From the aerial view around Ali astronomical observatory at A1 point (left panel of Figure \ref{fig:alisite}), we can see that, AliCPT site at B1 is not far from A1 point, the distance between them is only about 1km. The concrete road from A1 to B1 is already under construction and will be finished soon. In Ali astronomical observatory at A1 point, city grid electric power and the network infrastructure for data transmission are ready now. Some optical telescopes for astrophysics have been set up. Started in March 2017, the site construction of AliCPT is already ongoing and main building has been finished now. After the commissioning in 2019, the observation is expected to start in 2020.

\end{itemize}

In short, AliCPT opens a new window in the northern hemisphere to detect CMB B-mode polarization and probe the PGWs.

\begin{figure}
\includegraphics[scale=0.39]{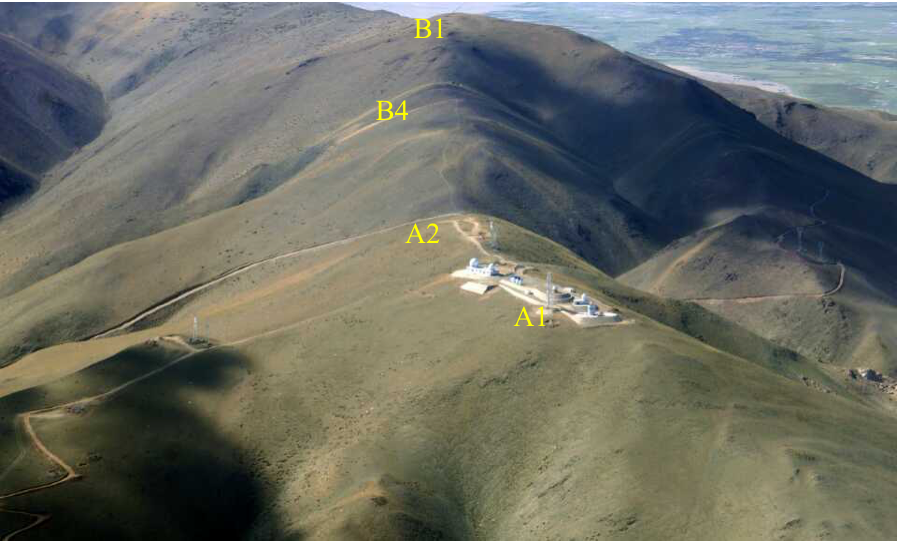}
\includegraphics[scale=0.114]{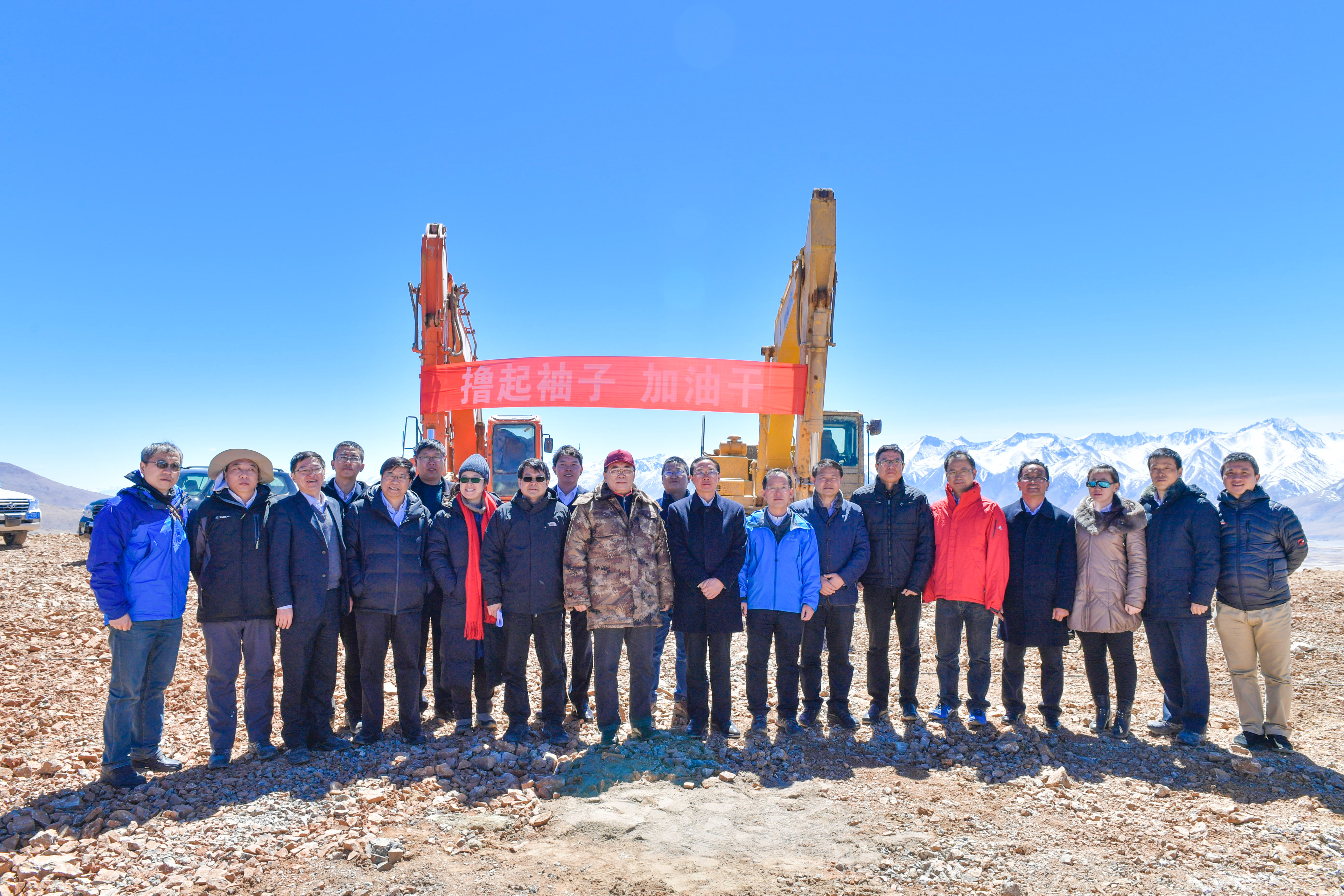}
\includegraphics[scale=0.25]{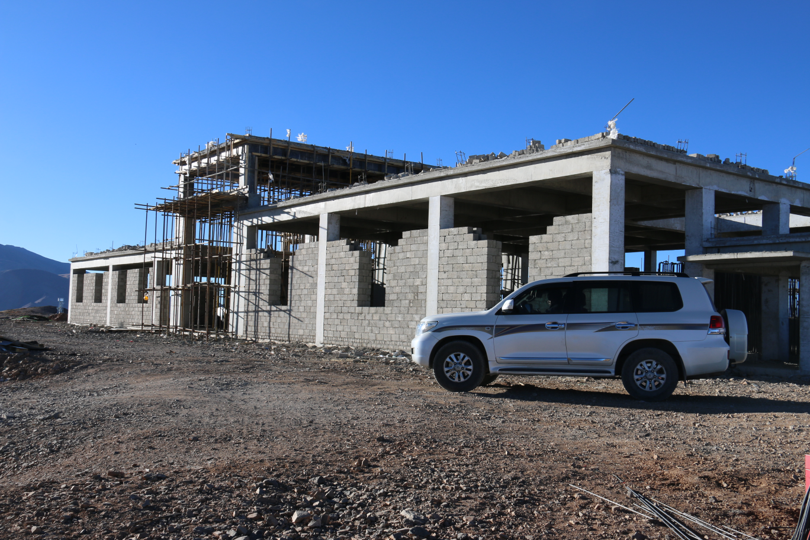}
	\caption{From left to right: aerial view of the region around Ali Astronomical Observatory (5100m at A1), picture for laying the foundation for AliCPT at B1 point (5250 m), the current infrastructure for AliCPT.
}
\label{fig:alisite}
\end{figure}

\section{AliCPT and its scientific goals}
\label{section3}

In this section, we provide a brief introduction to the main contents and scientific goals of the AliCPT project.
AliCPT project consists of two stages.
The first stage is to develop and deploy a CMB polarization telescope at 5,250 meters, called AliCPT-1.
AliCPT-1 telescope is a dichroic refractor of aperture 70cm covering 95/150GHz, with a three-axis driving mount scanning at speed of $5^{\circ}/s$ in azimuth.
AliCPT-1 adopts Transition Edge Sensor(TES) bolometers used widely in current CMB polarization experiments\cite{HeGao}, and takes Superconducting Quantum Interference Devices (SQUIDs) as cryogenic readout.
Sensors and their readout will be packaged into highly integrated modules, and each module contains 1,704 TES sensors.
The AliCPT-1 telescope will include four modules and the number of detectors reaches 6,816.

In the second stage, we will have a more sensitive telescope (AliCPT-2) with 12 modules and detectors more than 20,000.
The construction of AliCPT-2 will start in 2020. In each year, we will install four modules and finish the 12 modules by the end of 2022. In Table I, we summarize the basic instrumental parameters used for the simulations and the corresponding schedule of AliCPT-1 and AliCPT-2.

The scientific goals include:
\begin{itemize}
\item Start the large coverage surveys in northern hemisphere and search for regions with low foreground contamination;
\item Target at the cleanest sky region and detect PGWs in the northern hemisphere;
\item Measure the CMB rotation angle in high precision and test the CPT symmetry;
\item Study the hemispherical asymmetry in combination with experiments in southern hemisphere;
\item Measure E-mode polarization in high precision and study the effects in cosmology;
\item Study cross-correlation of CMB polarization with the large scale structures (LSS).
\end{itemize}
In the following, we will describe the sciences covered by AliCPT and provide our preliminary simulations on two main aspects of measurements on $r$ and the CMB rotation angle.

\begin{table}
\centering
\caption{Instrumental parameters for AliCPT-1 and AliCPT-2. Schedule for the number of modules and detectors installed. NET means the Noise Equivalent Temperature of each detector, $f_{\mathrm{sky}}$ represents sky coverage.}\label{inst}
\begin{tabular}{c|c|c|c|c}
\hline
\hline
      Year & 2019{\tiny{(AliCPT-1)}} & 2020{\tiny{(AliCPT-1+AliCPT-2)}}& 2021{\tiny{(AliCPT-1 + AliCPT-2)}}  & 2022\tiny{(AliCPT-1 + AliCPT-2)} \\
\hline
    $\mathrm{NET}(\mu K \sqrt{s})$     & 350 & 350 & 350 & 350\\
    $N_{\mathrm{\rm{mod}}}$ &  4   &  4+4  & 4+8 & 4+12\\
    $N_{\mathrm{det}}$ & 6,816 & 13,632 & 20,448 & 27,264 \\
    $f_{\mathrm{sky}}$ & $10\%$  & $10\%$  & $10\%$ & $10\%$ \\
    Bands(GHz) & 95 $\&$ 150 & 95 $\&$ 150 & 95 $\&$ 150 & 95 $\&$ 150   \\
\hline
\hline
\end{tabular}
\end{table}

\subsection{Sensitivity on $r$ and its implication on the Early Universe Physics}

The leading scenario of the early Universe is the inflationary cosmology, which describes an accelerating expanding phase occurred before radiation epoch.
Inflation resolves several conceptual issues of the Big Bang theory including the flatness, monopole, and horizon problems\cite{Guth:1981}.
Moreover, inflation explains the origin of primordial perturbations, with a mechanism that the quantum fluctuations of inflaton field were stretched to be classical perturbations by exponential expansion of space.
The primordial perturbations have three types: scalar, vector and tensor.
Scalar modes will eventually seed the CMB temperature anisotropies and also lead to the formation of large-scale structures in the Universe.
Tensor modes, which are dubbed as PGWs, will introduce CMB B-mode polarization, which is the target signal in AliCPT observations.
Conventionally, we often use $r$ to describe tensor perturbations, where $r$ means the ratio of the amplitudes of power spectra of primordial tensor $A_T$ and scalar modes $A_S$, via
\begin{align}
 r \equiv \frac{A_T}{A_S}~.
\end{align}

We have performed simulations to forecast the constraining capability of the AliCPT survey.
To obtain the constraint on $r$ derived from the AliCPT observations, we adopt the Fisher matrix approach \cite{Fisher}, which is an efficient way to forecast the constraints on cosmological parameters given the specification of the instruments.

Generally, the likelihood function of a multivariate Gaussian-distributed data vector $\mathbf{d}$ can be expressed as,
\begin{equation}
  \mathcal{L} = \frac{1}{\sqrt{|\mathbf{C(\theta)}|}} \exp\left(-\frac{1}{2}\mathbf{d^{\dagger}[\mathbf{C}(\theta)]^{-1}}\mathbf{d}\right),
\end{equation}
where $\theta$ is the parameter vector, and $\mathbf{C}$ is the covariance matrix which is a function of $\theta$ in general.
The Fisher matrix, $F_{ij}$, which is the second partial derivative of the likelihood function with respect to parameters $\theta_{i}$ and $\theta_{j}$ evaluated at the fiducial model, approximates the Hessian matrix, and the inverse of the diagonal terms $({\bf F}^{-1})_{ii}$ provides an estimate of the lower limit of the variance for parameter $\theta_{i}$, \ie, $\Delta\theta_i\gtrsim({\bf F}^{-1})_{ii}^{1/2}$ \cite{Fisher}.

For the case of CMB, $\mathbf{d} = \{X^{\nu}_{\ell m}, ...\}$, where X $\in\{T, E, B\}$, and $\nu$ runs over all available frequency bands.
Each $X^{\nu}_{\ell m}$ consists of three components, namely, the lensed CMB, the foreground emission and the instrumental noise.
The Fisher matrix for the CMB observarbles is
\begin{equation}\label{fisher}
  F_{ij} = \sum_{\ell}\frac{2\ell+1}{2}f_{\mathrm{sky}}\mathrm{Tr}\left[\mathbf{C}^{-1}_{\ell}\frac{\partial\mathbf{C}_{\ell}}{\partial\theta_{i}}\mathbf{C}^{-1}_{\ell}\frac{\partial\mathbf{C}_{\ell}}{\partial\theta_{j}}\right]~,
\end{equation}
where $\ell$ denotes the order of the multipole, $f_{\mathrm{sky}}$ is the sky coverage, and $\mathbf{C}_{\ell}$ is the covariance matrix in harmonic space which can be expressed as,
\begin{equation}
	\mathbf{C}_{\ell}(X^{\mu}_{\ell m}, Y^{\nu}_{\ell m}) = C^{XY, \mu\nu}_{\ell} + F^{XY, \mu\nu}_{\ell} + N^{XY, \mu\nu}_{\ell}~,
\end{equation}
where $C_{\ell}$, $F_{\ell}$, $N_{\ell}$ correspond to spectra of the lensed CMB, foreground and instrumental noise respectively.
We have assumed that observed temperature and polarization are statistically isotropic, so that terms of different $\ell$'s are independent.

\begin{figure}
  \centering
  \includegraphics[width=0.8\textwidth]{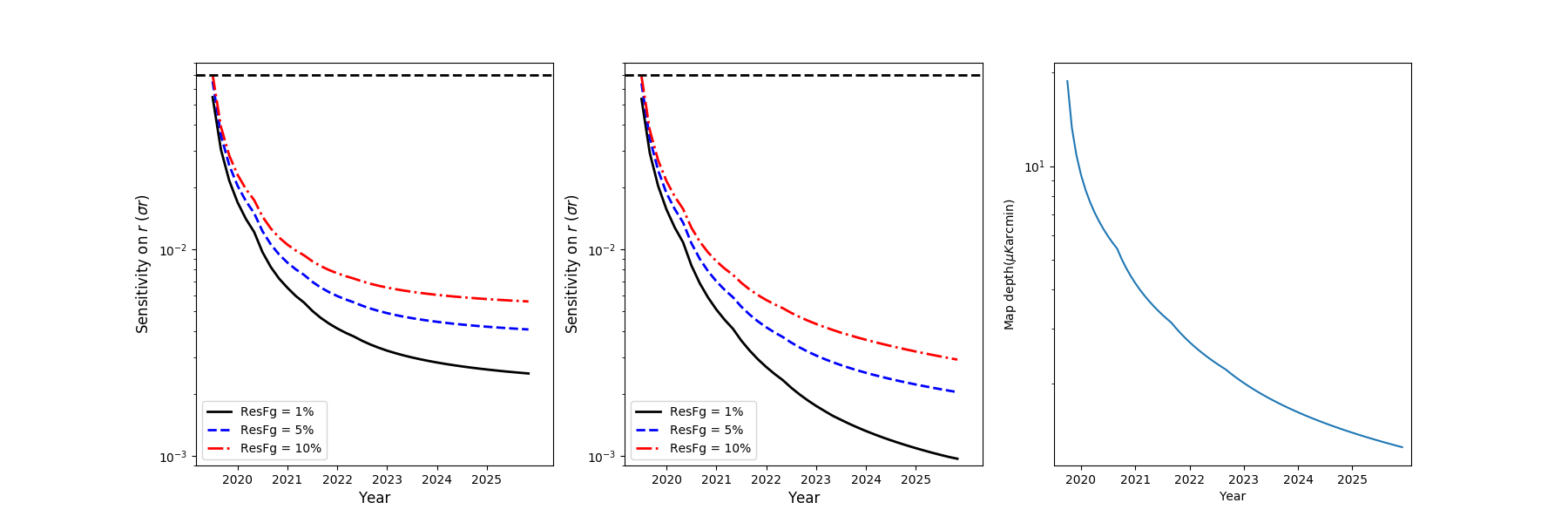}
  \caption{Left panel: Sensitivity of r without delensing. Middle panel: Sensitivity of r with completely delensing. The horizontal black dashed line is the current limit $r<0.07$. Right panel: Map depth of AliCPT.}
  \label{fig_r_cons}
\end{figure}

We use {\tt CAMB} \cite{CAMB} to calculate the CMB spectra $C_{\ell}$, and choose a fiducial cosmology which is consistent with the Planck 2015 results \cite{PLC15}.
Theoretically, the B-mode signal from lensing will exceed that from the primordial signal when on recombination bump ($\ell\sim 100$) $r$ falls below $0.01$.
Therefore, to achieve a high precision detection of the primordial B-mode signal, we shall perform a delensing procedure to remove the lensing effect from data.
To be remarkable, we consider two limit cases, totally-delensed and undelensed, in Figure \ref{fig_r_cons} for comparison.
During the actual operation, we will use the data of AliCPT itself, and its cross-correlation with the data of large scale surveys to reconstruct the lensing effect.
We will also attempt to explore the possibility of building a large aperture CMB telescope at Ali which will be helpful in delensing.

Based on results of recent experiments, foreground emission $F_{\ell}$ is thought to dominate over all frequency bands and all scales.
This kind of contamination can be removed through multi-frequency surveys, since the frequency spectrum distributions of CMB and other components are different.
Component separation is then defined as a method estimating the amount of each emission component.
Even such a separation has been applied, some extent of residual foreground still remains in data.
The AliCPT telescope plans to scan the cleanest sky regions at 95/150 GHz, in order to suppress residual foreground as much as possible.
In our Fisher forecasting, we consider two kinds of components, namely the synchrotron and thermal dust, which contribute to the dominate part of polarized foreground emission.
Besides $r$, we include six additional parameters in the free parameter space, which are $A_{\rm{dust}}$ and $A_{\rm{sync}}$ for amplitudes of dust and synchrotron at $\ell = 80$ with pivot frequencies of 353 GHz and 23 GHz, $\alpha_{\rm{dust}}$ and $\alpha_{\rm{sync}}$ for spectral index in harmonic space, $\beta_{\rm{dust}}$ and $\beta_{\rm{sync}}$ for spectral index in frequency space, respectively.
And no conversion between these two components, dust and synchrotron is assumed in our study.

In our simulations, we assume that the AliCPT instrumental noise of temperature and polarization are uncorrelated, and an isotropic Gaussian random noise for simplicity.
The noise spectrum is given by $N_{\ell}=w^{-1}B_{\ell}^2$ where $B_{\ell} = \exp(-\ell(\ell+1)\theta_{\mathrm{FWHM}}^2/8\log2)$ is the harmonic transform of the Gaussian beam
The weight is then,
\begin{equation}\label{noise}
  w^{-1} = \frac{4\pi f_{\mathrm{sky}}\mathrm{NET}^2}{t_{\mathrm{obs}}N_{\mathrm{det}}} ~,
\end{equation}
where NET means the Noise Equivalent Temperature which is closely related to the detector performance, instrument design and atmospheric condition, $t_{\mathrm{obs}}$ is the effective observation time (from October to March, 12h per day) and $N_{\mathrm{det}}$ is the number of detectors.
Note for polarization, NET should be multiplied by a factor of $\sqrt{2}$ since each polarized signal needs two orthogonal linear polarized detectors.
In Table \ref{inst}, we show the instrumental parameters for our calculation.

We plot the $1\sigma$ sensitivities of $r$ and map depth in Figure \ref{fig_r_cons}.
Results under undelensed and totally-delensed assumptions are showed in the left and middle panels.
Cases related to 1, 5, and 10 percent of residual foreground are considered in both panels.
Dashed black lines represent current representative constraint on $r$, $r<0.07$ at $2\sigma$, obtained from the joint analyses of BICEP2/Keck Array and Planck\cite{Array:2015xqh}.
As one can see from the plots, even in the undelensed case with 10 percent residual foreground, after 3 years' survey of AliCPT, the sensitivity on $r$ will reach $\sigma_{r} = 0.007$, which is one order of magnitude stronger than current constraint.
By the end of 2025, after 3 years' observation with 16 detector modules (around 27000 TES detectors), the constraint on $r$ in delensed case will reach a level of $0.003$.
Not surprisingly, less residual foreground leads to a better result on $r$.
If we can suppress foreground down to the one percent level, and apply a delensing procedure, we will obtain $\sigma_{r} \sim 0.001$.
Much stronger limit of $r$ can help us with testing a large number of cosmological models of the early Universe.
Due to its high precision, the AliCPT project will provide us a new vision of the early Universe.

\begin{figure}
\begin{center}
\includegraphics[scale=0.59]{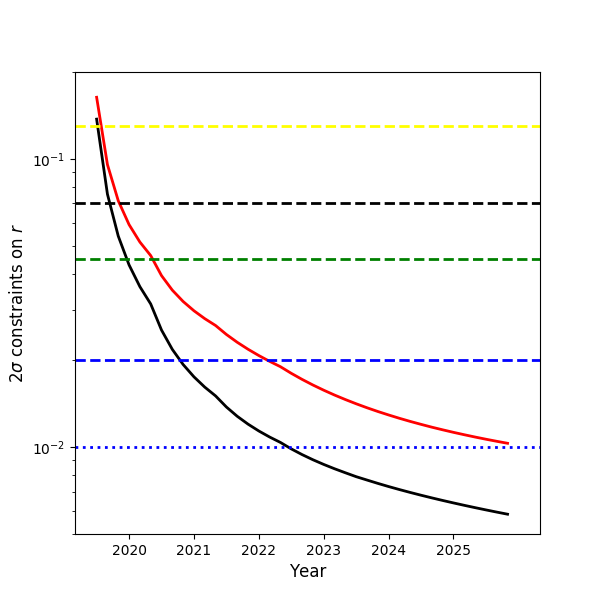}
\caption{Theoretical predictions of three inflation models and the scheduled AliCPT sensitivity of the measurements on $r$. The black and red curves represent the AliCPT $2\sigma$ limits on $r$, where in the simulations we have considered 30\% residual lensing effect, and residual foreground of 1\% (black) and 10\% (red). The black dashed line is the current limit from BICEP/Keck Array and Planck collaborations\cite{Array:2015xqh}.  The yellow dashed line, green dashed line, are the predictions from inflation models with potential function of $\phi^2$ \cite{phi2} and $\phi^{2/3}$ \cite{phi23}. The blue dashed and dotted lines are for the alpha attractor model \cite{alphaa} with $\alpha = 7$, $\alpha = 3$ and $n = 1$. In the theoretical calculations, e-folds number is taken to be 60.}
\label{fig:gw}
\end{center}
\end{figure}

In Figure \ref{fig:gw}, we plotted the scheduled $2\sigma$ constraints on $r$ and typical inflation model predictions, we can see that high precision measurements on PGWs are crucial in testing the inflation models.
Since inflationary cosmology still suffer from the initial cosmic singularity \cite{Penrose:1965, Borde:1994}, alternative theoretical approaches have been proposed, including bounce cosmology \cite{CaiRev:2014}, cyclic Universe \cite{LehnersRev:2008}, and emergent Universe \cite{Ellis:2004}.
Similar to inflation, these theories can also generate primordial tensor perturbations.
However, PGWs generated from different theories have different characteristics.
The precise measurement on the B-mode will help to distinguish among these models of early Universe.

\subsection{Sensitivity on the CMB Polarization Rotation Angle and its implication on the CPT Test}

Testing CPT symmetry, the combination of charge conjugation (C), parity reflection (P) and  time reversal (T) is important to cosmology and particle physics.
Any violation, if found, would be a powerful and important clue for new physics beyond the standard model.
So far, the laboratory experiments are consistent with a null result for the CPT violation. However, these tests may not be applied to physical processes in the early Universe at extremely high energy scales. In fact, there are motivations to speculate on the CPT violation in cosmology.
Firstly, the expanding Universe has a preferred temporal direction, which provides a natural frame to break the Lorentz and CPT symmetries.
Secondly, the observed baryon and anti-baryon asymmetry in the Universe may indicate a dynamical CPT violation \cite{Cohen:1988kt, Li:2001st}.

To study the cosmological CPT violation in CMB, we consider the following effective Lagrangian,
\be
\mathcal{L}_{CS}=p_{\mu}A_{\nu}\widetilde{F}^{\mu\nu}~,
\label{lag}
\ee
where the external field $p_{\mu}$ is a constant vector\cite{Carroll}, or $p_{\mu}\sim \partial_{\mu}\phi$ with $\phi$ being the dark energy scalar in quintessential baryo/leptogensis\cite{Li:2001st,Li:2006ss}, or $p_{\mu}\sim \partial_{\mu}R$ with $R$ the Ricci Scalar in gravitational baryo/leptogensis\cite{Davoudiasl:2004gf,Li:2004hh}, $\widetilde{F}^{\mu\nu}=(1/2)\epsilon^{\mu\nu\rho\sigma}F_{\rho\sigma}$ is the dual tensor of the electromagnetic tensor $F_{\mu\nu}$.
With the Chern-Simons term in (\ref{lag}), the polarization directions of photons get rotated for CMB, this will convert part of E-mode polarization to the B-mode and change the power spectra of polarization fields \cite{Lue:1998mq,Feng:2004mq,Feng:2006dp}. In Table \ref{tab:cpt}, we summarized the constraints on the rotation angle from various experiments. The current limit on the rotation angle $\alpha$ is about $1^\circ$.

\begin{table}
\caption {Measurements on CMB rotation angle since 2006. }\label{tab:cpt}
\begin{center}
   \begin{tabular}{ c | c | c }
   \hline			
    & ~~~Data~~~ & $\alpha+\sigma_\alpha^{stat}+\sigma_\alpha^{sys}$  \\
   \hline
  1&WMAP3+BOOMERANG \cite{Feng:2006dp} & $-6^{\circ}\pm4^{\circ}$\\
  2&WMAP3 \cite{paolo} & $-2.5^{\circ}\pm3.0^{\circ}$\\
  3&WMAP5 \cite{wmap5} & $-1.7^{\circ}\pm2.1^{\circ}$\\
  4&WMAP7 \cite{wmap7} & $-1.1^{\circ}\pm1.4^{\circ}\pm1.5^{\circ}$\\
  5&WMAP9 \cite{wmap9} & $-0.36^{\circ}\pm1.24^{\circ}\pm1.5^{\circ}$\\
  6&QUaD \cite{quad} & $-0.56^{\circ}\pm0.82^{\circ}\pm0.5^{\circ}$\\
  7&BICEP1 \cite{bicep1} & $-2.6^{\circ}\pm1.02^{\circ}$\\
  8&BICEP1 \cite{bicep1xia} & $-2.77^{\circ}\pm0.86^{\circ}\pm1.3^{\circ}$\\
  9&POLARBEAR \cite{polarbear} & $-1.08^{\circ}\pm0.20^{\circ}\pm0.5^{\circ}$\\
  10&ACTPol \cite{actpol} & $-0.2^{\circ}\pm0.5^{\circ}$\\
  11&Planck 2015 \cite{planckcpt} & $0.35^{\circ}\pm0.05^{\circ}\pm0.28^{\circ}$\\
  \hline
  \end{tabular}
\end{center}

\end{table}

We have performed simulations to forecast the sensitivity on the measurements of the rotation angle with AliCPTs.
For the forecast, we employ the so-called D-estimators\cite{quad} defined as follows,
\begin{eqnarray}
  D_{\ell}^{\mathrm{TB, obs}} &=&  C_{\ell}^{\mathrm{TB, obs}}\cos(2\beta) - C_{\ell}^{\mathrm{TE, obs}}\sin(2\beta);\nonumber\\
  D_{\ell}^{\mathrm{EB, obs}} &=&  C_{\ell}^{\mathrm{EB, obs}}\cos(4\beta) - \frac{1}{2}(C_{\ell}^{\mathrm{EE, obs}}-C_{\ell}^{\mathrm{BB, obs}})\sin(4\beta),
\end{eqnarray}
where $\beta$ is the unbiased estimator for the cosmic rotation angle $\alpha$.

\begin{figure}
  \centering
  \includegraphics[width=0.4\textwidth]{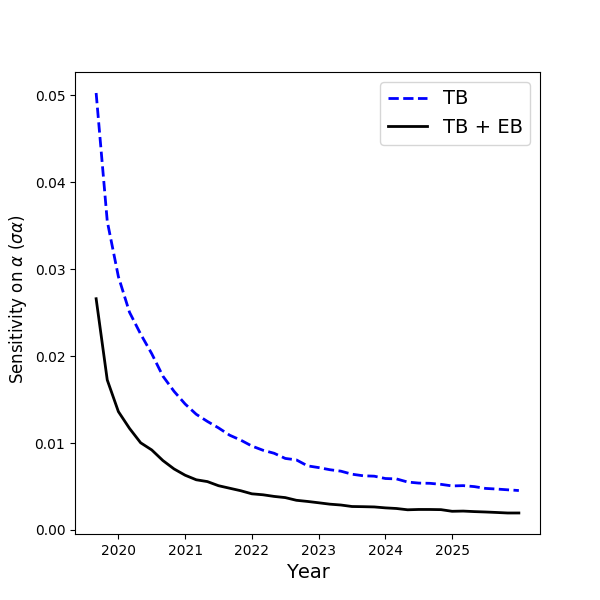}\\
  \caption{Forecast of average polarization rotation angle. The blue dashed line is obtained from TB estimator and the black solid line is from TB+EB estimator.}\label{fig_cpt}
\end{figure}

We modified the generic Monte Carlo Markov Chain (MCMC) sampler provided in the {\tt CosmoMC} package\cite{CosmoMC} and performed the calculation with instrumental properties of AliCPT.
We present our results in Figure \ref{fig_cpt}, where the blue dashed and black solid lines represent the constraints with TB and EB + TB estimator respectively.
Our results show that, after 3 years' observation, the AliCPTs are capable of providing a stringent constraint on the average rotation angle of $\sigma(\alpha) \sim 0.01^{\circ}$.\\

Before concluding, we mention briefly on the other topics listed in the beginning of this section.
One is the study on the hemispherical asymmetry of CMB polarization is also one important scientific goal of AliCPT.
After the Planck's survey, the temperature anisotropy has been precisely measured and well understood, which is basically found to be consistent with the standard cosmological model.
At the same time, however, some statistical analyses indicate that the CMB temperature suffer from a certain level of asymmetry.
For example, this kind of asymmetry can be directly modeled as a dipolar modulation through $\Delta T/T=s(\hat{n})[1+A\hat{n}\cdot\hat{p}]$, where $s(\hat{n})$ refers to a statistically isotropic CMB sky\cite{plkdipolar}.
The best-fit value is $A=0.072 \pm 0.022$ for CMB power at $l<64$, and pointing to $(227^{\circ},-27^{\circ})$.
There are some explanations to this phenomena, including the systematic errors, galactic contaminations and physically origins.
If the asymmetry is originated by a specific scenario such like the anisotropy of primordial fluctuations, a similar asymmetry pattern should also be observed in CMB polarization.
Therefore, measurement on the asymmetry of polarization maps is the key to distinguish various theories, and study the related physical mechanisms in the early Universe.
Combining with the CMB missions in the south, AliCPT will be helpful to realize the comparison between the polarization maps of different hemispheres, and will play an important role for investigating polarized hemispherical asymmetry.

In addition, the precise measurement on the CMB E-mode is crucial important to the studies on sciences related to the reionization.
Due to the large sky coverage, AliCPT will provide us an excellent opportunity to study the reionization history since the reionization bump corresponds to the CMB spectra with multipoles of $\ell < 10$.
Also, AliCPT plans to adopt the data of LSS tracers for the cross-correlating studies.
The high precision survey of DESI in the northern-hemisphere will be very much aligned with the AliCPT for the large overlapping scanning area.
Such cross-correlations will provide us a powerful way in resolving several topics like the reconstruction of ISW effect, CMB lensing, etc.

\section{Summary}
\label{section4}

In this paper, we have made an introduction to the new ground-based CMB polarization project in China, AliCPT, including the observational conditions and scientific goals.
The current AliCPT is located at a 5,250m hilltop in the Ali Prefecture of Tibet, and it is also planned to have another site nearby with an altitude of 6,000m for the forthcoming mission.
The median values of PWV crucial important to CMB experiments at 5,250m and 6,000m are around $1 mm$ and $0.6 mm$ from October to March. These are excellent to the observations of CMB polarization at 95/150GHz(5,250m), and higher frequencies(6,000m). Being in the mid-latitude, AliCPT will cover the whole northern hemisphere and partial southern hemisphere, and complement the CMB missions in the south, realizing a full sky coverage in the search of the PGWs.

The major scientific goals of AliCPT include searching for the PGWs and measuring the polarization rotation angle.
Our preliminary simulations show that, after three observing seasons, AliCPT will improve the current constraint on the tensor-to-scalar ratio by one order of magnitude and the current limit on rotation angle by two order of magnitude.


\textit{Acknowledgements.}
AliCPT project is supported in part by NSFC (Nos. 11653001, 11653002, 11653005, 11653004), CAS pilot B project of No. XDB23020000, and Sino US cooperation project No. 2016YFE0104700.

\end{document}